\documentclass[aps,prl,reprint,superscriptaddress]{revtex4-1}
\usepackage[dvipsnames]{xcolor}
\usepackage{graphicx}
\usepackage{amsmath,amssymb}
\usepackage{bm}
\usepackage{braket}
\usepackage{lipsum}
\usepackage{makerobust}
\usepackage{simpler-wick}
\usepackage{hyperref}

\newcommand*\circled[1]{\tikz[baseline=(char.base)]{
    \node[shape=circle,draw,inner sep=1pt] (char) {#1};}}
\MakeRobustCommand\circled

\newcommand{\makeauthor}[2]{\newcommand{#1}[1]{{%
  \sffamily\color{#2}{%
    \bfseries\begingroup\escapechar=-1\edef\x{\endgroup\string#1}\x:%
  } ##1}}%
  \MakeRobustCommand#1}
\makeauthor{\eric}{Plum}
\makeauthor{\themba}{ForestGreen} 
\makeauthor{\dc}{magenta}
\makeauthor{\sr}{black}
\makeauthor{\Fig}{red}

\begin{document}

\renewcommand{\vec}[1]{\bm{#1}}
\newcommand{\up}{{\uparrow}}
\newcommand{\dw}{{\downarrow}}
\newcommand{\pa}{{\partial}}
\newcommand{\pd}{{\phantom{\dagger}}}
\newcommand{\bs}[1]{\boldsymbol{#1}}
\newcommand{\add}[1]{{{\color{black}#1}}}
\newcommand{\todo}[1]{{\textbf{\color{red}ToDo: #1}}}
\newcommand{\tbr}[1]{{\textbf{\color{red}\underline{ToBeRemoved:} #1}}}
\newcommand{\eps}{{\varepsilon}}
\newcommand{\nn}{\nonumber}

\newcommand{\brap}[1]{{\bra{#1}_{\rm phys}}}
\newcommand{\bral}[1]{{\bra{#1}_{\rm log}}}
\newcommand{\ketp}[1]{{\ket{#1}_{\rm phys}}}
\newcommand{\ketl}[1]{{\ket{#1}_{\rm log}}}
\newcommand{\braketp}[1]{{\braket{#1}_{\rm phys}}}
\newcommand{\braketl}[1]{{\braket{#1}_{\rm log}}}

\graphicspath{{./}{./figures/}}


\title{Many-body Majorana braiding without an exponential Hilbert space}

\author{Eric Mascot}
\author{Themba Hodge}
\author{Dan Crawford}
\affiliation{School of Physics, University of Melbourne, Parkville, VIC 3010, Australia}
\author{Jasmin Bedow}
\author{Dirk K. Morr}
\affiliation{University of Illinois at Chicago, Chicago, IL 60607, USA}
\author{Stephan Rachel}
\affiliation{School of Physics, University of Melbourne, Parkville, VIC 3010, Australia}
\noaffiliation

\date{\today}

\begin{abstract}
Qubits built out of Majorana zero modes (MZMs) constitute the primary path towards topologically protected quantum computing. 
Simulating the braiding process of multiple MZMs corresponds to the quantum dynamics of a superconducting many-body system.
It is crucial to study the Majorana dynamics both in the presence of all other quasiparticles and for reasonably large system sizes. We present a method to calculate arbitrary many-body wavefunctions as well as their expectation values, correlators and overlaps from time evolved single-particle states of a superconductor, allowing for si\-gni\-fi\-cantly larger system sizes. We calculate the fidelity, transition probabilities, and joint parities of Majorana pairs to track the quality of the braiding process. We show how the braiding success depends on the speed of the braid. 
Moreover, we demonstrate the topological CNOT two-qubit gate as an example of two-qubit entanglement.
Our work opens the path to test and analyze the many theoretical implementations of Majorana qubits. Moreover, this method can be used to study the dynamics of any non-interacting superconductor.
\end{abstract}

\maketitle



{\it Introduction.---}
With the advent of commercially available noisy intermediate-scale quantum computers, the research focus has drifted towards {\it fault-tolerant} quantum computing concepts. Topological quantum computing might be the most exciting and fascinating strategy to accomplish fault-tolerance\,\cite{nayak-08rmp1083}. 
In a nutshell, exotic particles which obey non-Abelian braiding statistics are used to construct quantum bits (qubits). Unitary operations are realized by braiding of these anyons, and measurement is accomplished by analyzing the multi-anyon states. 
In the past decade topological superconductivity has proven to be a suitable platform for non-Abelian anyons\,\cite{read-00prb10267,kitaev2001}, due to the ability to artificially engineer such systems as nanostructures\,\cite{lutchyn-10prl077001,oreg-10prl177002}. Topological superconductors may host localized zero-energy subgap states, referred to as {\it Majorana zero modes} (MZMs), which are for practical purposes equivalent to Ising anyons\,\cite{markeshli-13prb045130,MZMvsIsing}. 
Today, MZMs are the most promising building blocks for future fault-tolerant quantum computers\,\cite{dassarma-15npjqi15001,lahtinen-17spp021,beenakker20sp15}. 

While the mathematical framework for topological quantum computing is well-established\,\cite{witten89cmp351,moore-91npb362,freedman-03bams31,nayak-96npb529}, it remains unclear what challenges and obstacles to expect when embedding this framework into a physical system. Realistic simulations are necessary to address these challenges and so we must contend with the fact that already one topological qubit consisting of four MZMs constitutes a superconducting quantum many-body system.
Exact diagonalization\,\cite{sekania2017,li2016,ED3} 
is most useful 
but 
limited to small system sizes. Many important contributions focus on single-particle states\,\cite{amorim2015,tutschku_majorana_2020} or study low-energy effective theories\,\cite{karzig_boosting_2013,chen-22prb054507,karzig_shortcuts_2015,knapp_the-nature_2016}, 
while ignoring bulk states and their potential effects on the MZMs.
A promising approach is to time-evolve the quasi-particles\,\cite{cheng2011,scheurer2013,sanno2021}, 
although simulation of key observables in a full many-body manner, such as transition amplitudes and parity measurements, have yet to be demonstrated. 
Majorana qubit errors have been studied using the Onishi formula\,\cite{conlon2019}, but this is subject to the sign problem \cite{mizusaki2018,robledo2009} for general use.
The covariance matrix of Refs.\,\cite{bravyi2012,bravyi2017,bauer-18sp004,truong2022} 
seems to have the potential to overcome most obstacles, although non-Abelian braiding has not yet been demonstrated.

%
%

In this Letter, we present a method for the efficient construction of superconducting ground and excited states from the single-particle basis. Time-evolving single-particle states and constructing many-body states, as known for Slater determinants, seem to be straight-forward. However, this task turns out to be significantly more challenging for a superconductor: (i) Time-evolving the many-body vacuum is non-trivial because this generates new quasi-particles, in contrast to the electronic case. That is, the vacuum state at different times is no longer proportional to itself.  (ii) Overlaps of many-body states lead to a Pfaffian with several anomalous blocks rather than a determinant with only one.
(iii) The biggest challenge is due to overlaps or expectation values at different times; because of (i) we end up with expressions where Wick's theorem cannot be applied. Our method resolves all these issues and thus  
avoids the exponentially large Hilbert space; that allows us to reach system sizes with more than a thousand lattice sites. 
We dynamically perform Pauli Z and X gate operations via braiding and show how the transition probabilities 
depend on the speed of the braid. 
We also present two-qubit entanglement by performing the topologically-protected CNOT gate.
Most notably, our method is applicable to the quantum dynamics of {\it any} superconducting many-body system.

%
%
{\it Method.---}We consider a general time-dependent Hamiltonian in the Bogoliubov-de Gennes (BdG) form,
\begin{equation}
    \mathcal{H}(t)
    =
    \frac{1}{2} \sum_{i, j}
    \begin{pmatrix} c_i^\dag & c_i^\pd \end{pmatrix}
    \begin{pmatrix} H_{ij}(t) & \Delta_{ij}(t) \\[5pt] \Delta_{ji}^*(t) & -H_{ij}^*(t) \end{pmatrix}
    \begin{pmatrix} c_j^\pd \\[5pt] c_j^\dag \end{pmatrix},
    \label{eq:HBdG}
\end{equation}
where $c_i^\dag$ 
creates 
an electron with index $i$, which can include site, momentum, orbital and spin.
$H(t)$ is the normal-state Hamiltonian matrix
and $\Delta(t)$ the pairing matrix.
The former 
is Hermitian, $H=H^\dag$, while the latter 
is anti-symmetric, $\Delta = -\Delta^T$.
We diagonalize the Hamiltonian at time $t=0$ with the Bogoliubov transformation,
\begin{equation}
    \begin{pmatrix} c_i^\pd \\[5pt] c_i^\dag \end{pmatrix}
    = \sum_n
    \begin{pmatrix} U_{in} & V_{in}^* \\[5pt] V_{in} & U_{in}^* \end{pmatrix}
    \begin{pmatrix} d_n^\pd \\[5pt] d_n^\dag \end{pmatrix},
    \label{eq:bogoliubov}
\end{equation}
which yields
\begin{equation}
    \mathcal{H}(0) = \sum_n E_n \left(
        d_n^\dag d_n^\pd - \frac{1}{2}
    \right).
    \label{eq:H0}
\end{equation}
We choose the energies, $E_n$, to be non-negative so that the quasiparticles, $d_n$, are excitations.
With this choice, it is clear that the ground state is the vacuum of quasiparticles, which we denote $\ket{\vec{0}_d}$, where $d_n \ket{\vec{0}_d} = 0$ for all $n$.
Such vacua are called Bogoliubov vacua.


One method to construct the Bogoliubov vacuum state is called the Thouless state, where $\ket{\vec{0}_d} \propto e^{\sum_{i, j} Z_{ij} c_i^\dag c_j^\dag} \ket{\vec{0}_c}$ and $Z = (V U^{-1})^*$\,\cite{shi2017}, which requires $U$ to be invertible.
Another method is called the product state\,\cite{shi2017}, where all quasiparticle operators are applied on the true vacuum, $\ket{\vec{0}_d} \propto \prod_n d_n \ket{\vec{0}_c}$.
This can, however, completely annihilate the vacuum, which occurs when $V$ is singular. To isolate the modes that annihilate the vacuum, we use the Bloch-Messiah decomposition\,\cite{bloch1962,ring1980,jin2022}.
The Bogoliubov matrix is decomposed as $U = C \bar{U} D^\dag$ and $V = C^* \bar{V} D^\dag$ where $C$ and $D$ are unitary and 
\begin{equation}
    \bar{U} = \begin{pmatrix}
        I & & \\
        & \oplus_k u_k \sigma_0 & \\
        & & 0
    \end{pmatrix},
    \quad
    \bar{V} = \begin{pmatrix}
        0 & & \\
        & \oplus_k v_k i\sigma_y & \\
        & & I
    \end{pmatrix},
    \label{eq:BM}
\end{equation}
with $u_k, v_k$ positive and $u_k^2 + v_k^2 = 1$.
The three blocks are called fully empty, paired, and fully occupied, corresponding to the zero, middle, and identity blocks of $\bar{V}$, 
respectively.
We define new operators
\begin{equation}
    c_i = \sum_j C_{ij} \bar{c}_j,
    \quad
    d_i = \sum_j D_{ij} \bar{d}_j.
    \label{eq:bar-basis}
\end{equation}
The $d$ and $\bar{d}$ quasiparticles share the same vacuum, which follows directly from Eq.\,\eqref{eq:bar-basis}.
We construct the product state using the $\bar{d}$ quasiparticles, truncating the fully empty modes:
\begin{eqnarray}
    \ket{\vec{0}_d} &=& \frac{1}{\sqrt{\mathcal{N}}} \prod_{k \in P} \bar{d}_k \bar{d}_{\bar{k}} \prod_{k \in O} \bar{d}_k \ket{\vec{0}_c} \\
    &=& \prod_{k \in P} \left(
        u_k + v_k \bar{c}_k^\dag \bar{c}_{\bar{k}}^\dag
    \right) \prod_{k \in O} \bar{c}_k^\dag \ket{\vec{0}_c}.
    \label{eq:d-vacuum}
\end{eqnarray}
$O$ denotes the fully occupied modes and $P$ denotes the paired modes.
The index, $k \in P$, iterates over paired indices, $(k, \bar{k})$, corresponding to the $2\times2$ blocks of $\bar{V}$.
The normalization is $\mathcal{N} = \prod_{k \in P} v_k^2$.

To construct excited states, we add excitations to the vacuum,
\begin{equation}
    \ket{\vec{n}_d} = \prod_k \left(d_k^\dag\right)^{n_k} \ket{\vec{0}_d},
    \label{eq:excited-state}
\end{equation}
where $n_k \in \{0, 1\}$ is the occupation of the $k$-th mode.
We then evolve the state with the time-evolution operator,
\begin{equation}
    \mathcal{S}(t) = \mathcal{T} \exp\left(
        -\frac{i}{\hbar} \int_0^t dt' \mathcal{H}(t')
    \right),
    \label{eq:time-evo-op}
\end{equation}
where $\mathcal{T}$ time-orders the exponential.
The time-evolved state is given by
\begin{equation}
    \ket{\vec{n}_d(t)} = \prod_k \left( d_k^\dag(t) \right)^{n_k} \ket{\vec{0}_d(t)},
\end{equation}
where $d_k^\dag(t) = \mathcal{S}(t) d_k^\dag \mathcal{S}^{-1}(t)$ and $\ket{\vec{0}_d(t)}$ is the vacuum of time-evolved quasiparticles, $d_k(t)$.

Time-evolving the operators is done using the time-dependent BdG equations \cite{cheng2011,amorim2015,conlon2019,sanno2021}, 
\begin{equation}
    i\hbar \frac{\partial}{\partial t}
    \begin{pmatrix}
        U(t) \\[3pt]
        V(t)
    \end{pmatrix}
    = H_\text{BdG}(t)
    \begin{pmatrix}
        U(t) \\[3pt]
        V(t)
    \end{pmatrix},
    \label{eq:TD-BdG}
\end{equation}
where $H_\text{BdG}(t)$ is the matrix in Eq.\,\eqref{eq:HBdG} and the initial conditions are given by the Bogoliubov transformation in Eq.\,\eqref{eq:bogoliubov}.
The time-evolved quasiparticles are given by
\begin{equation}
    \begin{pmatrix} c_i^\pd \\[5pt] c_i^\dag \end{pmatrix}
    = \sum_n
    \begin{pmatrix} U_{in}(t) & V_{in}^*(t) \\[5pt] V_{in}(t) & U_{in}^*(t) \end{pmatrix}
    \begin{pmatrix} d_n^\pd(t) \\[5pt] d_n^\dag(t) \end{pmatrix}.
    \label{eq:TD-qp}
\end{equation}
The time-evolved Bogoliubov vacuum is then given by
\begin{equation}
    \ket{\vec{0}_d(t)} = \frac{e^{i\varphi(t)}}{\sqrt{\mathcal{N}(t)}} \prod_{k \in P} \bar{d}_k(t) \bar{d}_{\bar{k}}(t) \prod_{k \in O} \bar{d}_k(t) \ket{\vec{0}_c},
    \label{eq:TD-d-vacuum}
\end{equation}
where $d_i(t) = \sum_j D_{ij}(t) \bar{d}_j(t)$ and $\mathcal{N}(t) = \prod_{k \in P} v_k^2(t)$, using the time-dependent Bloch-Messiah decomposition  $U(t) = C(t) \bar{U}(t) D^\dag(t)$ and $V(t) = C^*(t) \bar{V}(t) D^\dag(t)$.
$\varphi(t)$ is the phase difference between directly time-evolving the Bogoliubov vacuum and constructing the vacuum from time-evolved quasiparticles\,\cite{supp}.

\begin{table*}
\caption{\label{tab:contractions}
Contraction matrices, $M_{ij} = \braket{\vec{0}_c | a_i b_j | \vec{0}_c}$, between $a$ (row) and $b$ (column).}
\begin{ruledtabular}
\begin{tabular}{c|cccc}
&
$d^\dag(t')$ &
$d(t')$ &
$\bar{d}^\dag(t')$ &
$\bar{d}(t')$ \\ \hline
~~~~$d(t)$~~~~ &
$U^\dag(t) U(t')$ &
$U^\dag(t) V^*(t')$ &
$U^\dag(t) C(t') \bar{U}(t')$ &
$U^\dag(t) C(t') \bar{V}(t')$ \\
$d^\dag(t)$ &
$V^T(t) U(t')$ &
$V^T(t) V^*(t')$ &
$V^T(t) C(t') \bar{U}(t')$ &
$V^T(t) C(t') \bar{V}(t')$ \\
$\bar{d}(t)$ &
$\bar{U}(t) C^\dag(t) U(t')$ &
$\bar{U}(t) C^\dag(t) V^*(t')$ &
$\bar{U}(t) C^\dag(t) C(t') \bar{U}(t')$ &
$\bar{U}(t) C^\dag(t) C(t') \bar{V}(t')$ \\
$\bar{d}^\dag(t)$ &
$\bar{V}^T(t) C^\dag(t) U(t')$ &
$\bar{V}^T(t) C^\dag(t) V^*(t')$ &
$\bar{V}^T(t) C^\dag(t) C(t') \bar{U}(t')$ &
$\bar{V}^T(t) C^\dag(t) C(t') \bar{V}(t')$
\end{tabular}
\end{ruledtabular}
\end{table*}

Consider the quantity $\braket{\vec{n}_d(t) | \mathcal{A} | \vec{n}'_d(t')}$
where $\mathcal{A}$ is an arbitrary product of creation and annihilation operators.
For example, to calculate overlaps between states, we set $\mathcal{A} = 1$, or to calculate expectation values, we set $\ket{\vec{n}'_d(t')} = \ket{\vec{n}_d(t)}$.
By adapting a formula by Bertsch and Robledo\,\cite{bertsch2012,carlsson2021,jin2022} and extending it to the time-dependent case, this quantity becomes a Pfaffian:
%
\begin{equation}\begin{split}
   & \qquad\qquad \braket{\vec{n}_d(t) | \mathcal{A} | \vec{n}'_d(t')}
    = \pm \frac{e^{i(\varphi(t') - \varphi(t))}}{\sqrt{\mathcal{N}(t) \mathcal{N}(t')}} \times \\[5pt]
    &
    \text{pf} \begin{pmatrix}
        \wick{\c {\bar{d}}^\dag(t) \c {\bar{d}}^\dag(t)} &
        \wick{\c {\bar{d}}^\dag(t) \c d(t)} &
        \wick{\c {\bar{d}}^\dag(t) \c {\mathcal{A}}} &
        \wick{\c {\bar{d}}^\dag(t) \c d^\dag(t')} &
        \wick{\c {\bar{d}}^\dag(t) \c {\bar{d}}(t')} \\[3pt]
        &
        \wick{\c d(t) \c d(t)} &
        \wick{\c d(t) \c {\mathcal{A}}} &
        \wick{\c d(t) \c d^\dag(t')} &
        \wick{\c d(t) \c {\bar{d}}(t')} \\[3pt]
        &&
        \wick{\c {\mathcal{A}} \c {\mathcal{A}}} &
        \wick{\c {\mathcal{A}} \c d^\dag(t')} &
        \wick{\c {\mathcal{A}} \c {\bar{d}}(t')} \\[3pt]
        &&&
        \wick{\c d^\dag(t') \c d^\dag(t')} &
        \wick{\c d^\dag(t') \c {\bar{d}}(t')} \\[3pt]
        &&&&
        \wick{\c {\bar{d}}(t') \c {\bar{d}}(t')}
    \end{pmatrix}
\end{split}\end{equation}
The lower triangle is obtained by anti-symmetry.
The sign is derived from reversing the order of operators in $\bra{\vec{n}_d(t)}$ and is given by $\pm = (-1)^{n_{\bar{d}}(n_{\bar{d}}-1)/2 + n_d(n_d-1)/2}$ 
where $n_{\bar{d}}$ ($n_d$) is the number of $\bar{d}^\dag(t)$ ($d(t)$) operators in $\bra{\vec{n}_d(t)}$.
$\wick{\c a \c b}$ denotes the matrix of contractions such that $[\wick{\c a \c b}]_{ij} = \braket{\vec{0}_c|a_i b_j|\vec{0}_c}$.
This matrix only includes occupied modes and rows/columns of unoccupied modes are truncated.
We list useful contraction matrices in Table \ref{tab:contractions}. 
In the following, we will refer to the methodology used in this paper simply as the time-dependent Pfaffian (TDP) method.




%
%
{\it Results.---}For the remainder of the paper, we use the Kitaev chain\,\cite{kitaev2001} as the 
standard model of a topological superconductor hosting MZMs. 
It is given by Eq.\,\eqref{eq:HBdG} with $H_{ij}=-\mu_i(t)\delta_{ij} -\tilde{t} (\delta_{i,j+1} + \delta_{i+1,j})$ and $\Delta_{ij}=e^{i\phi}|\Delta_p| (\delta_{ij+1} - \delta_{j+1,i})$, where $\mu$, $\tilde{t}$, $\Delta_p$ and $\phi$ correspond to the chemical potential, tunneling strength, $p$-wave pairing strength and superconducting phase, respectively.
Throughout the paper we set $\tilde t=\Delta_p$; for details, see the supplemental material\,\cite{supp}.
The topological phase with MZMs at chain ends is realized for $-2\tilde{t} < \mu_{\rm topo} < 2\tilde{t}$
\,\cite{kitaev2001}. 



{\it Benchmarking: Time-Dependent Pfaffian Method vs.\ Exact Dia\-go\-nali\-zation.---}Our first result is the demonstration of a Pauli $Z$ gate on a T-junction.
The Kitaev chain model is mo\-di\-fied straight-forwardly to fit onto the T-junction so that all pairing phases on the horizontal (vertical) legs are $\phi=0$ ($\phi=\pi/2$), corresponding to a $p_x + i p_y$ superconductor\,\cite{supp}. 
To allow comparison between the TDP method and the full many-body states, we have chosen a leg length $L=5$, i.e., total number of sites $N=16$.
Time-evolution is approximated using the Krylov subspace method\,\cite{saad_analysis_1992,hochbruck_krylov_1997} for both exact diagonalization and TDP method.
In subsequent sections, we use a 4-th order implicit Runge-Kutta method\,\cite{supp}.

\begin{figure}[b!]
\centering
\includegraphics{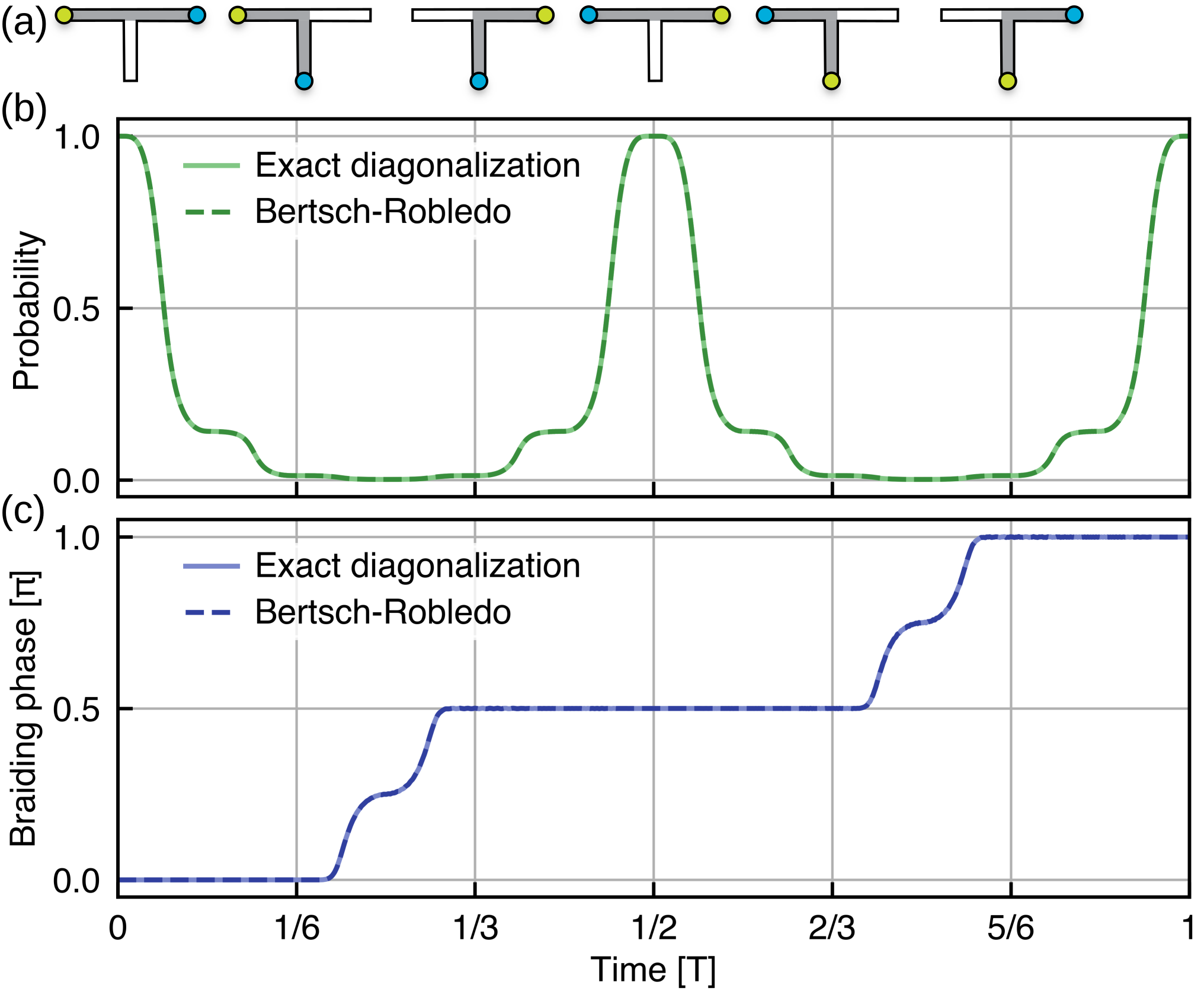}
\caption{Z gate on a T-junction ($L=5$). (a) Steps to perform braid. (b) Probability $|\braket{1|1(t)}_{\rm phys}|^2$
using both exact diagonalization and TDP method. (c) Braiding phase using both methods. Parameters: \((\mu_{\rm topo}, \mu_{\rm triv}, \alpha, T) = (-0.05\tilde{t}, -4\tilde{t}, 0.025, 960\hbar/\tilde{t})\).
 }
 \label{fig:bloch_messiah_vs_exact}
\end{figure}

The Z gate on the T-junction is realized by {\it moving} the Majorana modes via dynamical manipulation of the local chemical potential $\mu_i(t)$, as schematically illustrated in Fig.\,\ref{fig:bloch_messiah_vs_exact}\,a\,\cite{alicea2011}.
The total braiding time is denoted as $T$, and the delay coefficient $\alpha$ is the relative time between ramping the chemical potential of two neighboring sites.
That is, $\alpha=0$ means that $\mu_i(t)$ of all sites on a given leg are changed simultaneously, while $\alpha=1$ means that $\mu_i(t)$ of only a single site is changed at a time.
For details on the ramping protocol\,\cite{sekania2017} see the supplemental material\,\cite{supp}.
To guarantee the adiabaticity of the dynamics we consider the fidelities\,\cite{NielsenChuang2000} of the two degenerate many-body ground states $\ketp{0}$ and $\ketp{1}$. The former (latter) corresponds to the even (odd) parity sector (i.e., $\ketp{0}=\ket{\bs{0}_d}$), 
and they share the relationship $\ketp{1} = d^\dag \ketp{0}$, where for $\mu=0$ we simply have $d^\dag = \frac{1}{2}(\gamma_1 + i \gamma_{2N})$. The subscript ``phys'' refers to the physical Fock basis.

Fig.\,\ref{fig:bloch_messiah_vs_exact}\,b shows the squared fidelity or probability, $|\braket{1|1(t)}_{\rm phys}|^2$, which is here identical to $|\braket{0|0(t)}_{\rm phys}|^2$ (not shown).
Midway through the braid the fidelity returns to 1 even though the MZMs have been exchanged. 
That the states $\ketp{1}$ and $\ketp{1(T/2)}$ are nevertheless different is revealed through the braiding phase - the difference between the geometric phases\,\cite{mukunda1993,sekania2017} of the odd and even parity states - shown in Fig.\,\ref{fig:bloch_messiah_vs_exact}\,c. The geometric phase in its gauge- and parameterization-invariant form is $\phi_g(t)={\rm arg}\braket{\psi|\psi(t)}-{\rm Im}\int_0^t \braket{\psi(t')|\dot{\psi}(t')} dt'$.
After exchanging the MZMs once, a phase $\pi/2$ is accumulated, demonstrating the anyonic character of the MZMs and their fractional statistics, while realizing a $\sqrt{{\rm Z}}$ gate. After exchanging the MZMs once more, the MZMs return to their original position and the total braiding phase $\pi$ is reached, concluding the Z gate.
Moreover, the fidelities have returned to 1, ruling out any transitions into excited states.
The  many-body calculation of a Z gate can be found in the literature\,\cite{sekania2017}; here we only show it to highlight the agreement between exact diagonalization (solid lines) and the TDP method (dashed lines) in Fig.\,\ref{fig:bloch_messiah_vs_exact}\,b and c.

\begin{figure}[t!]
\centering
\includegraphics{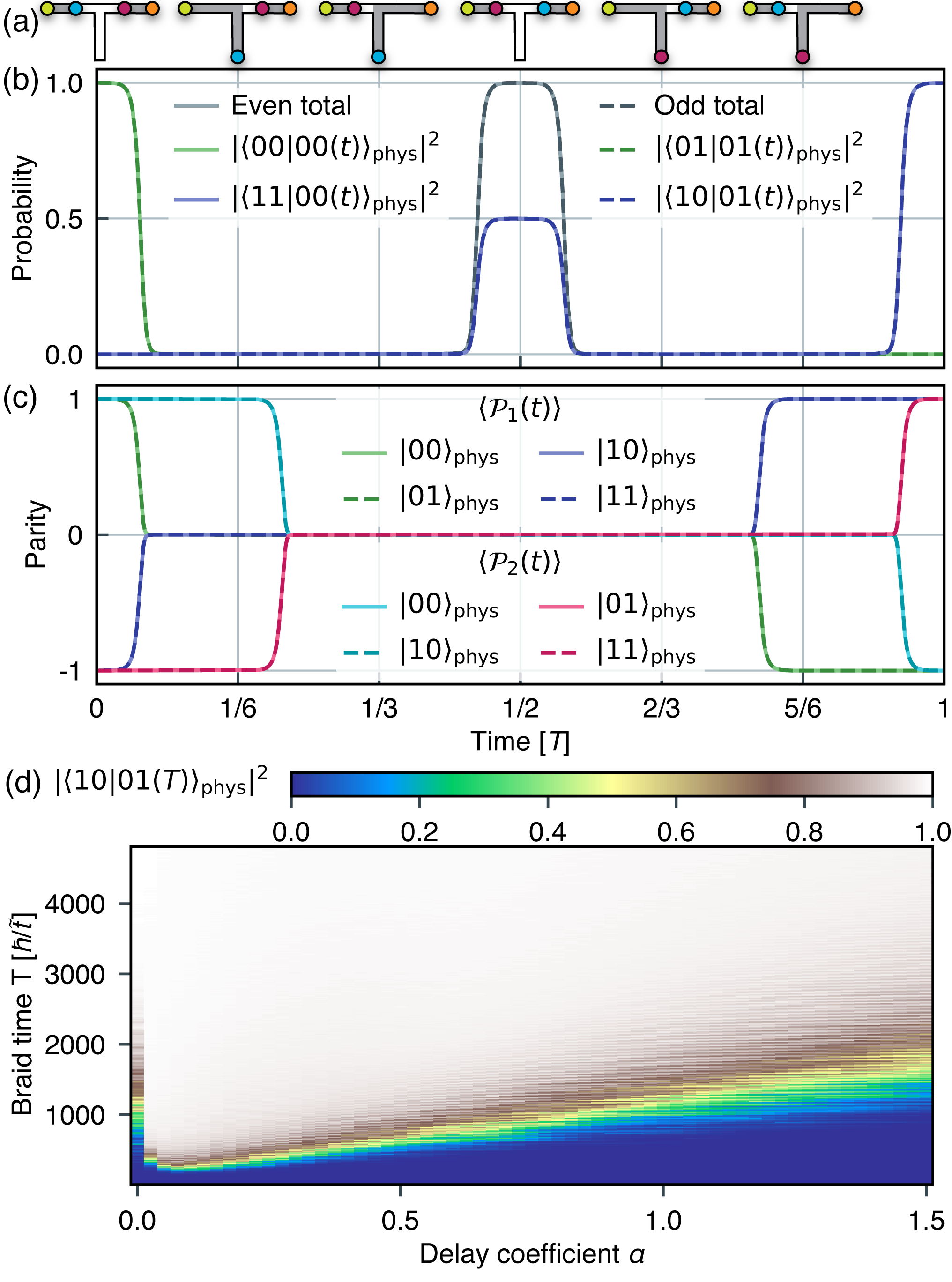} 
\caption{
X gate on a T-junction ($L=20$). (a) Steps to perform braid. (b) Transition probabilities. Green corresponds to same state transitions, with purple corresponding to different state transitions within a parity subspace. 
(c) Parities of the left MZM, $\braket{\mathcal{P}_1(t)}$, and right MZM, $\braket{\mathcal{P}_2(t)}$, for all 4 ground states.
(d) Transition probability $|\braket{10|01(T)}_{\rm phys}|^2$ as a function of braid time $T$ and delay coefficient $\alpha$.
Parameters: $(\mu_{\rm topo}, \mu_{\rm triv}, \alpha,  T)=(0.05\tilde{t}, 10\tilde{t}, 0.025, 15072\hbar/\tilde{t})$.}
 \label{fig:sigma-x_gate}
\end{figure}

{\it The X gate and non-Abelian braiding.---}
In the following we focus on a single qubit (i.e., four MZMs) and perform a Pauli X gate.
Fig.\,\ref{fig:sigma-x_gate}\,a illustrates the setup of two topological Kitaev chains on the T-junction; the X gate corresponds to exchanging the two inner MZMs twice. 
 In Fig.\,\ref{fig:sigma-x_gate}\,b we show the transition probabilities from one ground state $\ket{n}$ into the other one $\ket{m}$, $|\braket{n|m(t)}|^2$, and the ground state parities in Fig.\,\ref{fig:sigma-x_gate}\,c. 
 These ground states are either $\ketp{00}$ and $\ketp{11}$ (total parity = even) or $\ketp{10}$ and $\ketp{01}$ (total parity = odd)\,\cite{supp}, and simply correspond to the logical states $\ketl{0}$ and $\ketl{1}$ for either parity sector.
 At $t=0$, same-state probabilities are at unity and transition probabilities at zero. Midway through the braid, all transition probabilities reach 1/2 signifying that each initial state has transitioned to an equal superposition with the other initial state (corresponding to a $\sqrt{{\rm X}}$ gate), e.g. $\ketp{00} \to 
 (\ketp{00}-i\ketp{11})/\sqrt{2}$.
 Their total probability adds up to 1 indicating adiabaticity. Upon completion of the braid, the transition probability reaches 1 while the same-state probability goes to 0. We have successfully switched the Majorana qubit.

Our formalism allows to compute another useful quantity: the time-dependent parities of the original pairs of MZMs, i.e., of the two topological Kitaev chains, $\mathcal{P}_1$ and $\mathcal{P}_2$ with $\mathcal{P}_i = 1-2d_i^\dag d_i^\pd$ and $\braket{\mathcal{P}_i(t)}=\braket{\vec{n}(t) | \mathcal{P}_i | \vec{n}(t)}$. The total parity $\mathcal{P}_{\rm tot} = \braket{\mathcal{P}_1(t)} \braket{\mathcal{P}_2(t)}$ remains unchanged after the braid, but $\braket{\mathcal{P}_1(t)}$ and $\braket{\mathcal{P}_2(t)}$ will have swapped their initial values at the end of the braid. All eight combinations out of the two parity operators and the four ground states $\{ \ketp{00}, \ketp{11}, \ketp{10}, \ketp{01}\}$ are shown in Fig.\,\ref{fig:sigma-x_gate}\,c. 
At the end of the braid, all parities are opposite to their initial values.
The results confirm the transition probabilities of Fig.\,\ref{fig:sigma-x_gate}\,b. 
More broadly, we are able to compute arbitrary expectation values and correlation functions, as shown in the supplement\,\cite{supp}.

Fig.\,\ref{fig:sigma-x_gate}\,b also shows the total probabilities, e.g.\ $|\braket{00|00(t)}_{\rm phys}|^2+|\braket{11|00(t)}_{\rm phys}|^2$ in case of the even parity sector; at $t=0, T/2, T$ they reach unity, indicating the absence of transitions into excited states. That raises the question about robustness of such a braid upon varying braid time $T$ and delay coefficient $\alpha$. In Fig.\,\ref{fig:sigma-x_gate}\,d we present $|\braket{10|01(t)}_{\rm phys}|^2$ as a function of $T$ and $\alpha$. In the colorful regions, adiabaticity is violated and transitions out of the Majorana sector into excited states occur, spoiling the braid. Somewhat surprisingly, for rather short braid times of $T=1000 \hbar/\tilde{t}$ or $2000 \hbar/\tilde{t}$, the probability reaches already unity. Most remarkably, even at zero delay ($\alpha=0$), i.e., when the $\mu_i(t)$ are changed simultaneously for the entire leg, by slightly increasing $T$, even in this extreme case we find perfect transition probabilities. Assuming a hopping amplitude of the order of 0.1\,eV, we obtain e.g.\ a braid time $T=1000 \hbar/\tilde{t}\approx6.6$\,ps.

We have also repeated the X gate simulations on a T-junction embedded into a 2D substrate. By reaching system sizes as large as $N=1125$, we find that the results of Fig.\,\ref{fig:sigma-x_gate} remain unchanged\,\cite{supp}.

\begin{figure}[t!]
    \centering
    \includegraphics[width=\columnwidth]{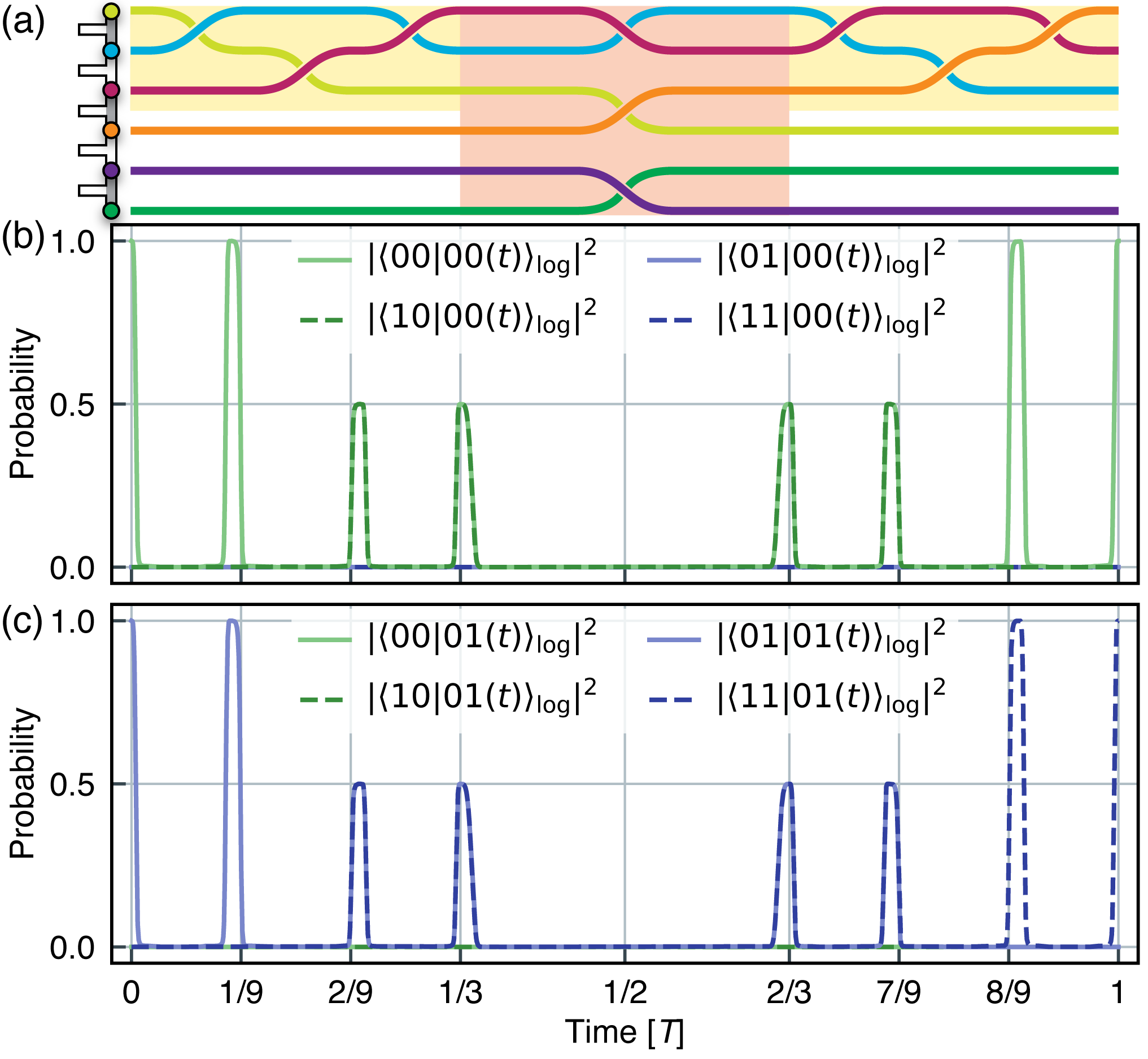}
    \caption{Controlled X gate (aka CNOT gate) on a multi-T junction geometry with six MZMs. (a) World lines of the MZMs, where the yellow boxes contain Hadamard gates on the first qubit, and the orange box the CZ gate. (b, c) Transition probabilities when the control qubit is (b) off and (c) on. Parameters: $(L, \mu_{\rm topo}, \mu_{\rm triv}, \alpha,  T)=(10, 0.05\tilde{t}, 10\tilde{t}, 0.025, 9217\hbar/\tilde{t})$.
    }
    \label{fig:CNOT}
\end{figure}

{\it The CNOT gate.---}One of the most important gates for quantum computing is the controlled X (CX) gate (aka CNOT gate).
As a two-qubit gate, it entangles neighboring qubits and is key to generating entangled many-qubit states.
It flips the first (target) qubit if and only if the second (control) qubit is in state $\ketl{1}$: ${\rm CX}\ketl{00}=\ketl{00}$, ${\rm CX}\ketl{01}=\ketl{11}$ etc.
We define three Kitaev chains with six MZMs at its ends on a multi-T junction in order to prepare two qubits for performing the CNOT gate, see Fig.\,\ref{fig:CNOT}\,a.
The CNOT gate is realizable using the identity $\rm CX = H_1 CZ H_1$ where CZ is the controlled Z gate (shown in orange in Fig.\,\ref{fig:CNOT}\,a) and $\rm H_1$ is the Hadamard gate on the target qubit (shown in yellow in Fig.\,\ref{fig:CNOT}\,a).
Details of the multi-T junction and braiding protocol along with the mapping from physical to logical qubit states can be found in the supplemental material\,\cite{supp}.
As expected for the CNOT gate, we find no flip at the end of the braid when the control qubit is off, $|\braket{00|00(T)}_{\rm log}|^2=1$ (see Fig.\,\ref{fig:CNOT}\,b); in contrast, the target qubit flips when the control qubit is on, $|\braket{11|01(T)}_{\rm log}|^2=1$ (see Fig.\,\ref{fig:CNOT}\,c).
This constitutes the explicit demonstration of two-qubit entanglement via braiding\,\cite{georgiev-06pra235112,georgiev-08npb552}.

{\it Discussion and Outlook.---}All results have been derived by using ``beneficial'' parameters to eliminate braiding errors. After having established that the general ideas of anyon braiding are indeed working in a dynamical many-body simulation, in the future one can focus on more realistic parameters and models \add{and explore the} role of disorder and electrostatic noise or low-lying subgap states (``quasi-particle poisoning''), \add{and test the limitations of a topological quantum computer}. The method reported in this Letter should help to answer many of these important questions; more broadly, it allows one to study the dynamics of an arbitrary superconducting many-body system.
Real-world problems of multi-Majorana systems under non-equilibrium conditions can be analyzed, as demonstrated here for Pauli Z and X gates as well as for the two-qubit CNOT gate.

By choosing a Majorana platform, one can obtain specific results, such as the braiding dynamics in magnet-superconductor hybrid systems\,\cite{bedow-23}. 
This method represents a powerful tool to perform numerical experiments, long before the real experiments are in reach.

The data and code for this Letter are openly available in Zenodo\,\cite{zenodo}.


\begin{acknowledgments}
We acknowledge discussions with Hong-Hao Tu, B. G. Carlsson and J. Rotureau.
S.R.\ acknowledges support from the Australian Research Council through Grant No.\ DP200101118.
J.B.\ and D.K.M.\ acknowledge support by the U.\ S.\ Department of Energy, Office of Science, Basic Energy Sciences, under Award No.\ DE-FG02-05ER46225. 
This research was undertaken using resources from the National Computational Infrastructure (NCI Australia), an NCRIS enabled capability supported by the Australian Government.
\end{acknowledgments}


\begin{thebibliography}{48}%
\makeatletter
\providecommand \@ifxundefined [1]{%
 \@ifx{#1\undefined}
}%
\providecommand \@ifnum [1]{%
 \ifnum #1\expandafter \@firstoftwo
 \else \expandafter \@secondoftwo
 \fi
}%
\providecommand \@ifx [1]{%
 \ifx #1\expandafter \@firstoftwo
 \else \expandafter \@secondoftwo
 \fi
}%
\providecommand \natexlab [1]{#1}%
\providecommand \enquote  [1]{``#1''}%
\providecommand \bibnamefont  [1]{#1}%
\providecommand \bibfnamefont [1]{#1}%
\providecommand \citenamefont [1]{#1}%
\providecommand \href@noop [0]{\@secondoftwo}%
\providecommand \href [0]{\begingroup \@sanitize@url \@href}%
\providecommand \@href[1]{\@@startlink{#1}\@@href}%
\providecommand \@@href[1]{\endgroup#1\@@endlink}%
\providecommand \@sanitize@url [0]{\catcode `\\12\catcode `\$12\catcode
  `\&12\catcode `\#12\catcode `\^12\catcode `\_12\catcode `\%12\relax}%
\providecommand \@@startlink[1]{}%
\providecommand \@@endlink[0]{}%
\providecommand \url  [0]{\begingroup\@sanitize@url \@url }%
\providecommand \@url [1]{\endgroup\@href {#1}{\urlprefix }}%
\providecommand \urlprefix  [0]{URL }%
\providecommand \Eprint [0]{\href }%
\providecommand \doibase [0]{http://dx.doi.org/}%
\providecommand \selectlanguage [0]{\@gobble}%
\providecommand \bibinfo  [0]{\@secondoftwo}%
\providecommand \bibfield  [0]{\@secondoftwo}%
\providecommand \translation [1]{[#1]}%
\providecommand \BibitemOpen [0]{}%
\providecommand \bibitemStop [0]{}%
\providecommand \bibitemNoStop [0]{.\EOS\space}%
\providecommand \EOS [0]{\spacefactor3000\relax}%
\providecommand \BibitemShut  [1]{\csname bibitem#1\endcsname}%
\let\auto@bib@innerbib\@empty
\bibitem [{\citenamefont {Nayak}\ \emph {et~al.}(2008)\citenamefont {Nayak},
  \citenamefont {Simon}, \citenamefont {Stern}, \citenamefont {Freedman},\ and\
  \citenamefont {Das~Sarma}}]{nayak-08rmp1083}%
  \BibitemOpen
  \bibfield  {author} {\bibinfo {author} {\bibfnamefont {C.}~\bibnamefont
  {Nayak}}, \bibinfo {author} {\bibfnamefont {S.~H.}\ \bibnamefont {Simon}},
  \bibinfo {author} {\bibfnamefont {A.}~\bibnamefont {Stern}}, \bibinfo
  {author} {\bibfnamefont {M.}~\bibnamefont {Freedman}}, \ and\ \bibinfo
  {author} {\bibfnamefont {S.}~\bibnamefont {Das~Sarma}},\ }\href {\doibase
  10.1103/RevModPhys.80.1083} {\bibfield  {journal} {\bibinfo  {journal} {Rev.
  Mod. Phys.}\ }\textbf {\bibinfo {volume} {80}},\ \bibinfo {pages} {1083}
  (\bibinfo {year} {2008})}\BibitemShut {NoStop}%
\bibitem [{\citenamefont {Read}\ and\ \citenamefont
  {Green}(2000)}]{read-00prb10267}%
  \BibitemOpen
  \bibfield  {author} {\bibinfo {author} {\bibfnamefont {N.}~\bibnamefont
  {Read}}\ and\ \bibinfo {author} {\bibfnamefont {D.}~\bibnamefont {Green}},\
  }\href {\doibase 10.1103/PhysRevB.61.10267} {\bibfield  {journal} {\bibinfo
  {journal} {Phys. Rev. B}\ }\textbf {\bibinfo {volume} {61}},\ \bibinfo
  {pages} {10267} (\bibinfo {year} {2000})}\BibitemShut {NoStop}%
\bibitem [{\citenamefont {Kitaev}(2001)}]{kitaev2001}%
  \BibitemOpen
  \bibfield  {author} {\bibinfo {author} {\bibfnamefont {A.~Y.}\ \bibnamefont
  {Kitaev}},\ }\href {\doibase 10.1070/1063-7869/44/10S/S29} {\bibfield
  {journal} {\bibinfo  {journal} {Phys.-Usp.}\ }\textbf {\bibinfo {volume}
  {44}},\ \bibinfo {pages} {131} (\bibinfo {year} {2001})}\BibitemShut
  {NoStop}%
\bibitem [{\citenamefont {Lutchyn}\ \emph {et~al.}(2010)\citenamefont
  {Lutchyn}, \citenamefont {Sau},\ and\ \citenamefont
  {Das~Sarma}}]{lutchyn-10prl077001}%
  \BibitemOpen
  \bibfield  {author} {\bibinfo {author} {\bibfnamefont {R.~M.}\ \bibnamefont
  {Lutchyn}}, \bibinfo {author} {\bibfnamefont {J.~D.}\ \bibnamefont {Sau}}, \
  and\ \bibinfo {author} {\bibfnamefont {S.}~\bibnamefont {Das~Sarma}},\ }\href
  {\doibase 10.1103/PhysRevLett.105.077001} {\bibfield  {journal} {\bibinfo
  {journal} {Phys. Rev. Lett.}\ }\textbf {\bibinfo {volume} {105}},\ \bibinfo
  {pages} {077001} (\bibinfo {year} {2010})}\BibitemShut {NoStop}%
\bibitem [{\citenamefont {Oreg}\ \emph {et~al.}(2010)\citenamefont {Oreg},
  \citenamefont {Refael},\ and\ \citenamefont {von Oppen}}]{oreg-10prl177002}%
  \BibitemOpen
  \bibfield  {author} {\bibinfo {author} {\bibfnamefont {Y.}~\bibnamefont
  {Oreg}}, \bibinfo {author} {\bibfnamefont {G.}~\bibnamefont {Refael}}, \ and\
  \bibinfo {author} {\bibfnamefont {F.}~\bibnamefont {von Oppen}},\ }\href
  {\doibase 10.1103/PhysRevLett.105.177002} {\bibfield  {journal} {\bibinfo
  {journal} {Phys. Rev. Lett.}\ }\textbf {\bibinfo {volume} {105}},\ \bibinfo
  {pages} {177002} (\bibinfo {year} {2010})}\BibitemShut {NoStop}%
\bibitem [{\citenamefont {Barkeshli}\ \emph {et~al.}(2013)\citenamefont
  {Barkeshli}, \citenamefont {Jian},\ and\ \citenamefont
  {Qi}}]{markeshli-13prb045130}%
  \BibitemOpen
  \bibfield  {author} {\bibinfo {author} {\bibfnamefont {M.}~\bibnamefont
  {Barkeshli}}, \bibinfo {author} {\bibfnamefont {C.-M.}\ \bibnamefont {Jian}},
  \ and\ \bibinfo {author} {\bibfnamefont {X.-L.}\ \bibnamefont {Qi}},\ }\href
  {\doibase 10.1103/PhysRevB.87.045130} {\bibfield  {journal} {\bibinfo
  {journal} {Phys. Rev. B}\ }\textbf {\bibinfo {volume} {87}},\ \bibinfo
  {pages} {045130} (\bibinfo {year} {2013})}\BibitemShut {NoStop}%
\bibitem [{MZM()}]{MZMvsIsing}%
  \BibitemOpen
  \href@noop {} {}\bibinfo {note} {Majorana zero modes possess $R$ matrices
  which are only projectively equivalent to those of Ising anyons.}\BibitemShut
  {Stop}%
\bibitem [{\citenamefont {Sarma}\ \emph {et~al.}(2015)\citenamefont {Sarma},
  \citenamefont {Freedman},\ and\ \citenamefont
  {Nayak}}]{dassarma-15npjqi15001}%
  \BibitemOpen
  \bibfield  {author} {\bibinfo {author} {\bibfnamefont {S.~D.}\ \bibnamefont
  {Sarma}}, \bibinfo {author} {\bibfnamefont {M.}~\bibnamefont {Freedman}}, \
  and\ \bibinfo {author} {\bibfnamefont {C.}~\bibnamefont {Nayak}},\
  }\href@noop {} {\bibfield  {journal} {\bibinfo  {journal} {npj Quant. Inf.}\
  }\textbf {\bibinfo {volume} {1}},\ \bibinfo {pages} {15001} (\bibinfo {year}
  {2015})}\BibitemShut {NoStop}%
\bibitem [{\citenamefont {Lahtinen}\ and\ \citenamefont
  {Pachos}(2017)}]{lahtinen-17spp021}%
  \BibitemOpen
  \bibfield  {author} {\bibinfo {author} {\bibfnamefont {V.}~\bibnamefont
  {Lahtinen}}\ and\ \bibinfo {author} {\bibfnamefont {J.~K.}\ \bibnamefont
  {Pachos}},\ }\href {\doibase 10.21468/SciPostPhys.3.3.021} {\bibfield
  {journal} {\bibinfo  {journal} {SciPost Phys.}\ }\textbf {\bibinfo {volume}
  {3}},\ \bibinfo {pages} {021} (\bibinfo {year} {2017})}\BibitemShut {NoStop}%
\bibitem [{\citenamefont {Beenakker}(2020)}]{beenakker20sp15}%
  \BibitemOpen
  \bibfield  {author} {\bibinfo {author} {\bibfnamefont {C.~W.~J.}\
  \bibnamefont {Beenakker}},\ }\href@noop {} {\bibfield  {journal} {\bibinfo
  {journal} {SciPost Phys. Lect. Notes}\ ,\ \bibinfo {pages} {15}} (\bibinfo
  {year} {2020})}\BibitemShut {NoStop}%
\bibitem [{\citenamefont {Witten}(1989)}]{witten89cmp351}%
  \BibitemOpen
  \bibfield  {author} {\bibinfo {author} {\bibfnamefont {E.}~\bibnamefont
  {Witten}},\ }\href@noop {} {\bibfield  {journal} {\bibinfo  {journal} {Comm.
  Math. Phys.}\ }\textbf {\bibinfo {volume} {121}},\ \bibinfo {pages} {351}
  (\bibinfo {year} {1989})}\BibitemShut {NoStop}%
\bibitem [{\citenamefont {Moore}\ and\ \citenamefont
  {Read}(1991)}]{moore-91npb362}%
  \BibitemOpen
  \bibfield  {author} {\bibinfo {author} {\bibfnamefont {G.}~\bibnamefont
  {Moore}}\ and\ \bibinfo {author} {\bibfnamefont {N.}~\bibnamefont {Read}},\
  }\href@noop {} {\bibfield  {journal} {\bibinfo  {journal} {Nucl. Phys. B}\
  }\textbf {\bibinfo {volume} {360}},\ \bibinfo {pages} {362} (\bibinfo {year}
  {1991})}\BibitemShut {NoStop}%
\bibitem [{\citenamefont {Freedman}\ \emph {et~al.}(2003)\citenamefont
  {Freedman}, \citenamefont {Kitaev}, \citenamefont {Larsen},\ and\
  \citenamefont {Wang}}]{freedman-03bams31}%
  \BibitemOpen
  \bibfield  {author} {\bibinfo {author} {\bibfnamefont {M.~H.}\ \bibnamefont
  {Freedman}}, \bibinfo {author} {\bibfnamefont {A.}~\bibnamefont {Kitaev}},
  \bibinfo {author} {\bibfnamefont {M.~J.}\ \bibnamefont {Larsen}}, \ and\
  \bibinfo {author} {\bibfnamefont {Z.}~\bibnamefont {Wang}},\ }\href@noop {}
  {\bibfield  {journal} {\bibinfo  {journal} {Bull. Amer. Math. Soc.}\ }\textbf
  {\bibinfo {volume} {40}},\ \bibinfo {pages} {31} (\bibinfo {year}
  {2003})}\BibitemShut {NoStop}%
\bibitem [{\citenamefont {Nayak}\ and\ \citenamefont
  {Wilczek}(1996)}]{nayak-96npb529}%
  \BibitemOpen
  \bibfield  {author} {\bibinfo {author} {\bibfnamefont {C.}~\bibnamefont
  {Nayak}}\ and\ \bibinfo {author} {\bibfnamefont {F.}~\bibnamefont
  {Wilczek}},\ }\href {\doibase https://doi.org/10.1016/0550-3213(96)00430-0}
  {\bibfield  {journal} {\bibinfo  {journal} {Nucl. Phys. B}\ }\textbf
  {\bibinfo {volume} {479}},\ \bibinfo {pages} {529} (\bibinfo {year}
  {1996})}\BibitemShut {NoStop}%
\bibitem [{\citenamefont {Sekania}\ \emph {et~al.}(2017)\citenamefont
  {Sekania}, \citenamefont {Plugge}, \citenamefont {Greiter}, \citenamefont
  {Thomale},\ and\ \citenamefont {Schmitteckert}}]{sekania2017}%
  \BibitemOpen
  \bibfield  {author} {\bibinfo {author} {\bibfnamefont {M.}~\bibnamefont
  {Sekania}}, \bibinfo {author} {\bibfnamefont {S.}~\bibnamefont {Plugge}},
  \bibinfo {author} {\bibfnamefont {M.}~\bibnamefont {Greiter}}, \bibinfo
  {author} {\bibfnamefont {R.}~\bibnamefont {Thomale}}, \ and\ \bibinfo
  {author} {\bibfnamefont {P.}~\bibnamefont {Schmitteckert}},\ }\href {\doibase
  10.1103/PhysRevB.96.094307} {\bibfield  {journal} {\bibinfo  {journal} {Phys.
  Rev. B}\ }\textbf {\bibinfo {volume} {96}},\ \bibinfo {pages} {094307}
  (\bibinfo {year} {2017})}\BibitemShut {NoStop}%
\bibitem [{\citenamefont {Li}\ \emph {et~al.}(2016)\citenamefont {Li},
  \citenamefont {Neupert}, \citenamefont {Bernevig},\ and\ \citenamefont
  {Yazdani}}]{li2016}%
  \BibitemOpen
  \bibfield  {author} {\bibinfo {author} {\bibfnamefont {J.}~\bibnamefont
  {Li}}, \bibinfo {author} {\bibfnamefont {T.}~\bibnamefont {Neupert}},
  \bibinfo {author} {\bibfnamefont {B.~A.}\ \bibnamefont {Bernevig}}, \ and\
  \bibinfo {author} {\bibfnamefont {A.}~\bibnamefont {Yazdani}},\ }\href
  {\doibase 10.1038/ncomms10395} {\bibfield  {journal} {\bibinfo  {journal}
  {Nat. Commun.}\ }\textbf {\bibinfo {volume} {7}},\ \bibinfo {pages} {10395}
  (\bibinfo {year} {2016})}\BibitemShut {NoStop}%
\bibitem [{\citenamefont {Wieckowski}\ \emph {et~al.}(2020)\citenamefont
  {Wieckowski}, \citenamefont {Mierzejewski},\ and\ \citenamefont
  {Kupczynski}}]{ED3}%
  \BibitemOpen
  \bibfield  {author} {\bibinfo {author} {\bibfnamefont {A.}~\bibnamefont
  {Wieckowski}}, \bibinfo {author} {\bibfnamefont {M.}~\bibnamefont
  {Mierzejewski}}, \ and\ \bibinfo {author} {\bibfnamefont {M.}~\bibnamefont
  {Kupczynski}},\ }\href {\doibase 10.1103/PhysRevB.101.014504} {\bibfield
  {journal} {\bibinfo  {journal} {Phys. Rev. B}\ }\textbf {\bibinfo {volume}
  {101}},\ \bibinfo {pages} {014504} (\bibinfo {year} {2020})}\BibitemShut
  {NoStop}%
\bibitem [{\citenamefont {Amorim}\ \emph {et~al.}(2015)\citenamefont {Amorim},
  \citenamefont {Ebihara}, \citenamefont {Yamakage}, \citenamefont {Tanaka},\
  and\ \citenamefont {Sato}}]{amorim2015}%
  \BibitemOpen
  \bibfield  {author} {\bibinfo {author} {\bibfnamefont {C.~S.}\ \bibnamefont
  {Amorim}}, \bibinfo {author} {\bibfnamefont {K.}~\bibnamefont {Ebihara}},
  \bibinfo {author} {\bibfnamefont {A.}~\bibnamefont {Yamakage}}, \bibinfo
  {author} {\bibfnamefont {Y.}~\bibnamefont {Tanaka}}, \ and\ \bibinfo {author}
  {\bibfnamefont {M.}~\bibnamefont {Sato}},\ }\href {\doibase
  10.1103/PhysRevB.91.174305} {\bibfield  {journal} {\bibinfo  {journal} {Phys.
  Rev. B}\ }\textbf {\bibinfo {volume} {91}},\ \bibinfo {pages} {174305}
  (\bibinfo {year} {2015})}\BibitemShut {NoStop}%
\bibitem [{\citenamefont {Tutschku}\ \emph {et~al.}(2020)\citenamefont
  {Tutschku}, \citenamefont {Reinthaler}, \citenamefont {Lei}, \citenamefont
  {MacDonald},\ and\ \citenamefont {Hankiewicz}}]{tutschku_majorana_2020}%
  \BibitemOpen
  \bibfield  {author} {\bibinfo {author} {\bibfnamefont {C.}~\bibnamefont
  {Tutschku}}, \bibinfo {author} {\bibfnamefont {R.~W.}\ \bibnamefont
  {Reinthaler}}, \bibinfo {author} {\bibfnamefont {C.}~\bibnamefont {Lei}},
  \bibinfo {author} {\bibfnamefont {A.~H.}\ \bibnamefont {MacDonald}}, \ and\
  \bibinfo {author} {\bibfnamefont {E.~M.}\ \bibnamefont {Hankiewicz}},\ }\href
  {\doibase 10.1103/PhysRevB.102.125407} {\bibfield  {journal} {\bibinfo
  {journal} {Phys. Rev. B}\ }\textbf {\bibinfo {volume} {102}},\ \bibinfo
  {pages} {125407} (\bibinfo {year} {2020})}\BibitemShut {NoStop}%
\bibitem [{\citenamefont {Karzig}\ \emph {et~al.}(2013)\citenamefont {Karzig},
  \citenamefont {Refael},\ and\ \citenamefont {von
  Oppen}}]{karzig_boosting_2013}%
  \BibitemOpen
  \bibfield  {author} {\bibinfo {author} {\bibfnamefont {T.}~\bibnamefont
  {Karzig}}, \bibinfo {author} {\bibfnamefont {G.}~\bibnamefont {Refael}}, \
  and\ \bibinfo {author} {\bibfnamefont {F.}~\bibnamefont {von Oppen}},\ }\href
  {\doibase 10.1103/PhysRevX.3.041017} {\bibfield  {journal} {\bibinfo
  {journal} {Phys. Rev. X}\ }\textbf {\bibinfo {volume} {3}},\ \bibinfo {pages}
  {041017} (\bibinfo {year} {2013})}\BibitemShut {NoStop}%
\bibitem [{\citenamefont {Chen}\ \emph {et~al.}(2022)\citenamefont {Chen},
  \citenamefont {Wang}, \citenamefont {Wu}, \citenamefont {Qi}, \citenamefont
  {Liu},\ and\ \citenamefont {Xie}}]{chen-22prb054507}%
  \BibitemOpen
  \bibfield  {author} {\bibinfo {author} {\bibfnamefont {W.}~\bibnamefont
  {Chen}}, \bibinfo {author} {\bibfnamefont {J.}~\bibnamefont {Wang}}, \bibinfo
  {author} {\bibfnamefont {Y.}~\bibnamefont {Wu}}, \bibinfo {author}
  {\bibfnamefont {J.}~\bibnamefont {Qi}}, \bibinfo {author} {\bibfnamefont
  {J.}~\bibnamefont {Liu}}, \ and\ \bibinfo {author} {\bibfnamefont {X.~C.}\
  \bibnamefont {Xie}},\ }\href {\doibase 10.1103/PhysRevB.105.054507}
  {\bibfield  {journal} {\bibinfo  {journal} {Phys. Rev. B}\ }\textbf {\bibinfo
  {volume} {105}},\ \bibinfo {pages} {054507} (\bibinfo {year}
  {2022})}\BibitemShut {NoStop}%
\bibitem [{\citenamefont {Karzig}\ \emph {et~al.}(2015)\citenamefont {Karzig},
  \citenamefont {Pientka}, \citenamefont {Refael},\ and\ \citenamefont {von
  Oppen}}]{karzig_shortcuts_2015}%
  \BibitemOpen
  \bibfield  {author} {\bibinfo {author} {\bibfnamefont {T.}~\bibnamefont
  {Karzig}}, \bibinfo {author} {\bibfnamefont {F.}~\bibnamefont {Pientka}},
  \bibinfo {author} {\bibfnamefont {G.}~\bibnamefont {Refael}}, \ and\ \bibinfo
  {author} {\bibfnamefont {F.}~\bibnamefont {von Oppen}},\ }\href {\doibase
  10.1103/PhysRevB.91.201102} {\bibfield  {journal} {\bibinfo  {journal} {Phys.
  Rev. B}\ }\textbf {\bibinfo {volume} {91}},\ \bibinfo {pages} {201102}
  (\bibinfo {year} {2015})}\BibitemShut {NoStop}%
\bibitem [{\citenamefont {Knapp}\ \emph {et~al.}(2016)\citenamefont {Knapp},
  \citenamefont {Zaletel}, \citenamefont {Liu}, \citenamefont {Cheng},
  \citenamefont {Bonderson},\ and\ \citenamefont
  {Nayak}}]{knapp_the-nature_2016}%
  \BibitemOpen
  \bibfield  {author} {\bibinfo {author} {\bibfnamefont {C.}~\bibnamefont
  {Knapp}}, \bibinfo {author} {\bibfnamefont {M.}~\bibnamefont {Zaletel}},
  \bibinfo {author} {\bibfnamefont {D.~E.}\ \bibnamefont {Liu}}, \bibinfo
  {author} {\bibfnamefont {M.}~\bibnamefont {Cheng}}, \bibinfo {author}
  {\bibfnamefont {P.}~\bibnamefont {Bonderson}}, \ and\ \bibinfo {author}
  {\bibfnamefont {C.}~\bibnamefont {Nayak}},\ }\href {\doibase
  10.1103/PhysRevX.6.041003} {\bibfield  {journal} {\bibinfo  {journal} {Phys.
  Rev. X}\ }\textbf {\bibinfo {volume} {6}},\ \bibinfo {pages} {041003}
  (\bibinfo {year} {2016})}\BibitemShut {NoStop}%
\bibitem [{\citenamefont {Cheng}\ \emph {et~al.}(2011)\citenamefont {Cheng},
  \citenamefont {Galitski},\ and\ \citenamefont {Das~Sarma}}]{cheng2011}%
  \BibitemOpen
  \bibfield  {author} {\bibinfo {author} {\bibfnamefont {M.}~\bibnamefont
  {Cheng}}, \bibinfo {author} {\bibfnamefont {V.}~\bibnamefont {Galitski}}, \
  and\ \bibinfo {author} {\bibfnamefont {S.}~\bibnamefont {Das~Sarma}},\ }\href
  {\doibase 10.1103/PhysRevB.84.104529} {\bibfield  {journal} {\bibinfo
  {journal} {Phys. Rev. B}\ }\textbf {\bibinfo {volume} {84}},\ \bibinfo
  {pages} {104529} (\bibinfo {year} {2011})}\BibitemShut {NoStop}%
\bibitem [{\citenamefont {Scheurer}\ and\ \citenamefont
  {Shnirman}(2013)}]{scheurer2013}%
  \BibitemOpen
  \bibfield  {author} {\bibinfo {author} {\bibfnamefont {M.~S.}\ \bibnamefont
  {Scheurer}}\ and\ \bibinfo {author} {\bibfnamefont {A.}~\bibnamefont
  {Shnirman}},\ }\href {\doibase 10.1103/PhysRevB.88.064515} {\bibfield
  {journal} {\bibinfo  {journal} {Phys. Rev. B}\ }\textbf {\bibinfo {volume}
  {88}},\ \bibinfo {pages} {064515} (\bibinfo {year} {2013})}\BibitemShut
  {NoStop}%
\bibitem [{\citenamefont {Sanno}\ \emph {et~al.}(2021)\citenamefont {Sanno},
  \citenamefont {Miyazaki}, \citenamefont {Mizushima},\ and\ \citenamefont
  {Fujimoto}}]{sanno2021}%
  \BibitemOpen
  \bibfield  {author} {\bibinfo {author} {\bibfnamefont {T.}~\bibnamefont
  {Sanno}}, \bibinfo {author} {\bibfnamefont {S.}~\bibnamefont {Miyazaki}},
  \bibinfo {author} {\bibfnamefont {T.}~\bibnamefont {Mizushima}}, \ and\
  \bibinfo {author} {\bibfnamefont {S.}~\bibnamefont {Fujimoto}},\ }\href
  {\doibase 10.1103/PhysRevB.103.054504} {\bibfield  {journal} {\bibinfo
  {journal} {Phys. Rev. B}\ }\textbf {\bibinfo {volume} {103}},\ \bibinfo
  {pages} {054504} (\bibinfo {year} {2021})}\BibitemShut {NoStop}%
\bibitem [{\citenamefont {Conlon}\ \emph {et~al.}(2019)\citenamefont {Conlon},
  \citenamefont {Pellegrino}, \citenamefont {Slingerland}, \citenamefont
  {Dooley},\ and\ \citenamefont {Kells}}]{conlon2019}%
  \BibitemOpen
  \bibfield  {author} {\bibinfo {author} {\bibfnamefont {A.}~\bibnamefont
  {Conlon}}, \bibinfo {author} {\bibfnamefont {D.}~\bibnamefont {Pellegrino}},
  \bibinfo {author} {\bibfnamefont {J.~K.}\ \bibnamefont {Slingerland}},
  \bibinfo {author} {\bibfnamefont {S.}~\bibnamefont {Dooley}}, \ and\ \bibinfo
  {author} {\bibfnamefont {G.}~\bibnamefont {Kells}},\ }\href {\doibase
  10.1103/PhysRevB.100.134307} {\bibfield  {journal} {\bibinfo  {journal}
  {Phys. Rev. B}\ }\textbf {\bibinfo {volume} {100}},\ \bibinfo {pages}
  {134307} (\bibinfo {year} {2019})}\BibitemShut {NoStop}%
\bibitem [{\citenamefont {Mizusaki}\ \emph {et~al.}(2018)\citenamefont
  {Mizusaki}, \citenamefont {Oi},\ and\ \citenamefont
  {Shimizu}}]{mizusaki2018}%
  \BibitemOpen
  \bibfield  {author} {\bibinfo {author} {\bibfnamefont {T.}~\bibnamefont
  {Mizusaki}}, \bibinfo {author} {\bibfnamefont {M.}~\bibnamefont {Oi}}, \ and\
  \bibinfo {author} {\bibfnamefont {N.}~\bibnamefont {Shimizu}},\ }\href
  {\doibase 10.1016/j.physletb.2018.02.012} {\bibfield  {journal} {\bibinfo
  {journal} {Phys. Lett. B}\ }\textbf {\bibinfo {volume} {779}},\ \bibinfo
  {pages} {237} (\bibinfo {year} {2018})}\BibitemShut {NoStop}%
\bibitem [{\citenamefont {Robledo}(2009)}]{robledo2009}%
  \BibitemOpen
  \bibfield  {author} {\bibinfo {author} {\bibfnamefont {L.~M.}\ \bibnamefont
  {Robledo}},\ }\href {\doibase 10.1103/PhysRevC.79.021302} {\bibfield
  {journal} {\bibinfo  {journal} {Phys. Rev. C}\ }\textbf {\bibinfo {volume}
  {79}},\ \bibinfo {pages} {021302} (\bibinfo {year} {2009})}\BibitemShut
  {NoStop}%
\bibitem [{\citenamefont {Bravyi}\ and\ \citenamefont
  {K{\"o}nig}(2012)}]{bravyi2012}%
  \BibitemOpen
  \bibfield  {author} {\bibinfo {author} {\bibfnamefont {S.}~\bibnamefont
  {Bravyi}}\ and\ \bibinfo {author} {\bibfnamefont {R.}~\bibnamefont
  {K{\"o}nig}},\ }\href {\doibase 10.1007/s00220-012-1606-9} {\bibfield
  {journal} {\bibinfo  {journal} {Commun. Math. Phys.}\ }\textbf {\bibinfo
  {volume} {316}},\ \bibinfo {pages} {641} (\bibinfo {year}
  {2012})}\BibitemShut {NoStop}%
\bibitem [{\citenamefont {Bravyi}\ and\ \citenamefont
  {Gosset}(2017)}]{bravyi2017}%
  \BibitemOpen
  \bibfield  {author} {\bibinfo {author} {\bibfnamefont {S.}~\bibnamefont
  {Bravyi}}\ and\ \bibinfo {author} {\bibfnamefont {D.}~\bibnamefont
  {Gosset}},\ }\href {\doibase 10.1007/s00220-017-2976-9} {\bibfield  {journal}
  {\bibinfo  {journal} {Commun. Math. Phys.}\ }\textbf {\bibinfo {volume}
  {356}},\ \bibinfo {pages} {451} (\bibinfo {year} {2017})}\BibitemShut
  {NoStop}%
\bibitem [{\citenamefont {Bauer}\ \emph {et~al.}(2018)\citenamefont {Bauer},
  \citenamefont {Karzig}, \citenamefont {Mishmash}, \citenamefont {Antipov},\
  and\ \citenamefont {Alicea}}]{bauer-18sp004}%
  \BibitemOpen
  \bibfield  {author} {\bibinfo {author} {\bibfnamefont {B.}~\bibnamefont
  {Bauer}}, \bibinfo {author} {\bibfnamefont {T.}~\bibnamefont {Karzig}},
  \bibinfo {author} {\bibfnamefont {R.~V.}\ \bibnamefont {Mishmash}}, \bibinfo
  {author} {\bibfnamefont {A.~E.}\ \bibnamefont {Antipov}}, \ and\ \bibinfo
  {author} {\bibfnamefont {J.}~\bibnamefont {Alicea}},\ }\href {\doibase
  10.21468/SciPostPhys.5.1.004} {\bibfield  {journal} {\bibinfo  {journal}
  {SciPost Phys.}\ }\textbf {\bibinfo {volume} {5}},\ \bibinfo {pages} {004}
  (\bibinfo {year} {2018})}\BibitemShut {NoStop}%
\bibitem [{\citenamefont {Truong}\ \emph {et~al.}(2023)\citenamefont {Truong},
  \citenamefont {Agarwal},\ and\ \citenamefont {Pereg-Barnea}}]{truong2022}%
  \BibitemOpen
  \bibfield  {author} {\bibinfo {author} {\bibfnamefont {B.~P.}\ \bibnamefont
  {Truong}}, \bibinfo {author} {\bibfnamefont {K.}~\bibnamefont {Agarwal}}, \
  and\ \bibinfo {author} {\bibfnamefont {T.}~\bibnamefont {Pereg-Barnea}},\
  }\href {\doibase 10.1103/PhysRevB.107.104516} {\bibfield  {journal} {\bibinfo
   {journal} {Phys. Rev. B}\ }\textbf {\bibinfo {volume} {107}},\ \bibinfo
  {pages} {104516} (\bibinfo {year} {2023})}\BibitemShut {NoStop}%
\bibitem [{\citenamefont {Shi}\ and\ \citenamefont {Zhang}(2017)}]{shi2017}%
  \BibitemOpen
  \bibfield  {author} {\bibinfo {author} {\bibfnamefont {H.}~\bibnamefont
  {Shi}}\ and\ \bibinfo {author} {\bibfnamefont {S.}~\bibnamefont {Zhang}},\
  }\href {\doibase 10.1103/PhysRevB.95.045144} {\bibfield  {journal} {\bibinfo
  {journal} {Phys. Rev. B}\ }\textbf {\bibinfo {volume} {95}},\ \bibinfo
  {pages} {045144} (\bibinfo {year} {2017})}\BibitemShut {NoStop}%
\bibitem [{\citenamefont {Bloch}\ and\ \citenamefont
  {Messiah}(1962)}]{bloch1962}%
  \BibitemOpen
  \bibfield  {author} {\bibinfo {author} {\bibfnamefont {C.}~\bibnamefont
  {Bloch}}\ and\ \bibinfo {author} {\bibfnamefont {A.}~\bibnamefont
  {Messiah}},\ }\href {\doibase 10.1016/0029-5582(62)90377-2} {\bibfield
  {journal} {\bibinfo  {journal} {Nucl. Phys.}\ }\textbf {\bibinfo {volume}
  {39}},\ \bibinfo {pages} {95} (\bibinfo {year} {1962})}\BibitemShut {NoStop}%
\bibitem [{\citenamefont {Ring}\ and\ \citenamefont {Schuck}(1980)}]{ring1980}%
  \BibitemOpen
  \bibfield  {author} {\bibinfo {author} {\bibfnamefont {P.}~\bibnamefont
  {Ring}}\ and\ \bibinfo {author} {\bibfnamefont {P.}~\bibnamefont {Schuck}},\
  }\href@noop {} {\emph {\bibinfo {title} {The Nuclear Many-Body Problem}}},\
  \bibinfo {edition} {1st}\ ed.,\ \bibinfo {number} {1864-5879}\ (\bibinfo
  {publisher} {{Springer Berlin, Heidelberg}},\ \bibinfo {year}
  {1980})\BibitemShut {NoStop}%
\bibitem [{\citenamefont {Jin}\ \emph {et~al.}(2022)\citenamefont {Jin},
  \citenamefont {Sun}, \citenamefont {Zhou},\ and\ \citenamefont
  {Tu}}]{jin2022}%
  \BibitemOpen
  \bibfield  {author} {\bibinfo {author} {\bibfnamefont {H.-K.}\ \bibnamefont
  {Jin}}, \bibinfo {author} {\bibfnamefont {R.-Y.}\ \bibnamefont {Sun}},
  \bibinfo {author} {\bibfnamefont {Y.}~\bibnamefont {Zhou}}, \ and\ \bibinfo
  {author} {\bibfnamefont {H.-H.}\ \bibnamefont {Tu}},\ }\href {\doibase
  10.1103/PhysRevB.105.L081101} {\bibfield  {journal} {\bibinfo  {journal}
  {Phys. Rev. B}\ }\textbf {\bibinfo {volume} {105}},\ \bibinfo {pages}
  {L081101} (\bibinfo {year} {2022})}\BibitemShut {NoStop}%
\bibitem [{sup()}]{supp}%
  \BibitemOpen
  \href@noop {} {}\bibinfo {note} {See Supplemental Material at
  http://link.aps.org/supplemental/10.1103/... for details about the model,
  method, ramping protocol and supporting results as well as LDOS-animations
  for $\sqrt{\rm Z}$ and $\sqrt{\rm X}$ gates, and X gate on a
  substrate, which includes Refs.~\cite{wimmer2012,balian1969,hara1979,onishi1966,lieb1968,caianiello1973,terhal2002,hairer1993}}\BibitemShut {NoStop}%
\bibitem [{\citenamefont {Bertsch}\ and\ \citenamefont
  {Robledo}(2012)}]{bertsch2012}%
  \BibitemOpen
  \bibfield  {author} {\bibinfo {author} {\bibfnamefont {G.~F.}\ \bibnamefont
  {Bertsch}}\ and\ \bibinfo {author} {\bibfnamefont {L.~M.}\ \bibnamefont
  {Robledo}},\ }\href {\doibase 10.1103/PhysRevLett.108.042505} {\bibfield
  {journal} {\bibinfo  {journal} {Phys. Rev. Lett.}\ }\textbf {\bibinfo
  {volume} {108}},\ \bibinfo {pages} {042505} (\bibinfo {year}
  {2012})}\BibitemShut {NoStop}%
\bibitem [{\citenamefont {Carlsson}\ and\ \citenamefont
  {Rotureau}(2021)}]{carlsson2021}%
  \BibitemOpen
  \bibfield  {author} {\bibinfo {author} {\bibfnamefont {B.~G.}\ \bibnamefont
  {Carlsson}}\ and\ \bibinfo {author} {\bibfnamefont {J.}~\bibnamefont
  {Rotureau}},\ }\href {\doibase 10.1103/PhysRevLett.126.172501} {\bibfield
  {journal} {\bibinfo  {journal} {Phys. Rev. Lett.}\ }\textbf {\bibinfo
  {volume} {126}},\ \bibinfo {pages} {172501} (\bibinfo {year}
  {2021})}\BibitemShut {NoStop}%
\bibitem [{\citenamefont {Saad}(1992)}]{saad_analysis_1992}%
  \BibitemOpen
  \bibfield  {author} {\bibinfo {author} {\bibfnamefont {Y.}~\bibnamefont
  {Saad}},\ }\href {\doibase 10.1137/0729014} {\bibfield  {journal} {\bibinfo
  {journal} {SIAM J. on Numer. Anal.}\ }\textbf {\bibinfo {volume} {29}}
  (\bibinfo {year} {1992}),\ 10.1137/0729014}\BibitemShut {NoStop}%
\bibitem [{\citenamefont {Hochbruck}\ and\ \citenamefont
  {Lubich}(1997)}]{hochbruck_krylov_1997}%
  \BibitemOpen
  \bibfield  {author} {\bibinfo {author} {\bibfnamefont {M.}~\bibnamefont
  {Hochbruck}}\ and\ \bibinfo {author} {\bibfnamefont {C.}~\bibnamefont
  {Lubich}},\ }\href {\doibase 10.1137/S0036142995280572} {\bibfield  {journal}
  {\bibinfo  {journal} {SIAM J. Numer. Anal.}\ }\textbf {\bibinfo {volume}
  {34}},\ \bibinfo {pages} {1911} (\bibinfo {year} {1997})}\BibitemShut
  {NoStop}%
\bibitem [{\citenamefont {Alicea}\ \emph {et~al.}(2011)\citenamefont {Alicea},
  \citenamefont {Oreg}, \citenamefont {Refael}, \citenamefont {{von Oppen}},\
  and\ \citenamefont {Fisher}}]{alicea2011}%
  \BibitemOpen
  \bibfield  {author} {\bibinfo {author} {\bibfnamefont {J.}~\bibnamefont
  {Alicea}}, \bibinfo {author} {\bibfnamefont {Y.}~\bibnamefont {Oreg}},
  \bibinfo {author} {\bibfnamefont {G.}~\bibnamefont {Refael}}, \bibinfo
  {author} {\bibfnamefont {F.}~\bibnamefont {{von Oppen}}}, \ and\ \bibinfo
  {author} {\bibfnamefont {M.~P.~A.}\ \bibnamefont {Fisher}},\ }\href {\doibase
  10.1038/nphys1915} {\bibfield  {journal} {\bibinfo  {journal} {Nat. Phys.}\
  }\textbf {\bibinfo {volume} {7}},\ \bibinfo {pages} {412} (\bibinfo {year}
  {2011})}\BibitemShut {NoStop}%
\bibitem [{\citenamefont {Nielsen}\ and\ \citenamefont
  {Chuang}(2000)}]{NielsenChuang2000}%
  \BibitemOpen
  \bibfield  {author} {\bibinfo {author} {\bibfnamefont {M.~A.}\ \bibnamefont
  {Nielsen}}\ and\ \bibinfo {author} {\bibfnamefont {I.~L.}\ \bibnamefont
  {Chuang}},\ }\href@noop {} {\emph {\bibinfo {title} {Quantum Computation and
  Quantum Information}}}\ (\bibinfo  {publisher} {Cambridge University Press,
  UK},\ \bibinfo {year} {2000})\BibitemShut {NoStop}%
\bibitem [{\citenamefont {Mukunda}\ and\ \citenamefont
  {Simon}(1993)}]{mukunda1993}%
  \BibitemOpen
  \bibfield  {author} {\bibinfo {author} {\bibfnamefont {N.}~\bibnamefont
  {Mukunda}}\ and\ \bibinfo {author} {\bibfnamefont {R.}~\bibnamefont
  {Simon}},\ }\href {\doibase 10.1006/aphy.1993.1093} {\bibfield  {journal}
  {\bibinfo  {journal} {Ann. Phys.}\ }\textbf {\bibinfo {volume} {228}},\
  \bibinfo {pages} {205} (\bibinfo {year} {1993})}\BibitemShut {NoStop}%
\bibitem [{\citenamefont {Georgiev}(2006)}]{georgiev-06pra235112}%
  \BibitemOpen
  \bibfield  {author} {\bibinfo {author} {\bibfnamefont {L.~S.}\ \bibnamefont
  {Georgiev}},\ }\href {\doibase 10.1103/PhysRevB.74.235112} {\bibfield
  {journal} {\bibinfo  {journal} {Phys. Rev. B}\ }\textbf {\bibinfo {volume}
  {74}},\ \bibinfo {pages} {235112} (\bibinfo {year} {2006})}\BibitemShut
  {NoStop}%
\bibitem [{\citenamefont {Georgiev}(2008)}]{georgiev-08npb552}%
  \BibitemOpen
  \bibfield  {author} {\bibinfo {author} {\bibfnamefont {L.~S.}\ \bibnamefont
  {Georgiev}},\ }\href {\doibase 10.1016/j.nuclphysb.2007.07.016} {\bibfield
  {journal} {\bibinfo  {journal} {Nucl. Phys. B}\ }\textbf {\bibinfo {volume}
  {789}},\ \bibinfo {pages} {552} (\bibinfo {year} {2008})}\BibitemShut
  {NoStop}%
\bibitem [{\citenamefont {Bedow}\ \emph {et~al.}()\citenamefont {Bedow},
  \citenamefont {Mascot}, \citenamefont {Hodge}, \citenamefont {Rachel},\ and\
  \citenamefont {Morr}}]{bedow-23}%
  \BibitemOpen
  \bibfield  {author} {\bibinfo {author} {\bibfnamefont {J.}~\bibnamefont
  {Bedow}}, \bibinfo {author} {\bibfnamefont {E.}~\bibnamefont {Mascot}},
  \bibinfo {author} {\bibfnamefont {T.}~\bibnamefont {Hodge}}, \bibinfo
  {author} {\bibfnamefont {S.}~\bibnamefont {Rachel}}, \ and\ \bibinfo {author}
  {\bibfnamefont {D.~K.}\ \bibnamefont {Morr}},\ }\href@noop {} {\enquote
  {\bibinfo {title} {Implementation of topological quantum gates in
  magnet-superconductor hybrid structures},}\ }\bibinfo {note}
  {{a}rXiv:2302.04889}\BibitemShut {NoStop}%
\bibitem [{\citenamefont {Wimmer}(2012)}]{wimmer2012}%
  \BibitemOpen
  \bibfield  {author} {\bibinfo {author} {\bibfnamefont {M.}~\bibnamefont
  {Wimmer}},\ }\bibfield  {title} {\bibinfo {title} {Algorithm 923: Efficient
  numerical computation of the {Pfaffian} for dense and banded skew-symmetric
  matrices},\ }\href {https://doi.org/10.1145/2331130.2331138} {\bibfield
  {journal} {\bibinfo  {journal} {ACM Trans. Math. Softw.}\ }\textbf {\bibinfo
  {volume} {38}},\ \bibinfo {pages} {30:1} (\bibinfo {year}
  {2012})}\BibitemShut {NoStop}%
\bibitem [{\citenamefont {Balian}\ and\ \citenamefont
  {Brezin}(1969)}]{balian1969}%
  \BibitemOpen
  \bibfield  {author} {\bibinfo {author} {\bibfnamefont {R.}~\bibnamefont
  {Balian}}\ and\ \bibinfo {author} {\bibfnamefont {E.}~\bibnamefont
  {Brezin}},\ }\bibfield  {title} {\bibinfo {title} {Nonunitary bogoliubov
  transformations and extension of {{Wick}}'s theorem},\ }\href
  {https://doi.org/10.1007/BF02710281} {\bibfield  {journal} {\bibinfo
  {journal} {Nuovo Cimento B (1965-1970)}\ }\textbf {\bibinfo {volume} {64}},\
  \bibinfo {pages} {37} (\bibinfo {year} {1969})}\BibitemShut {NoStop}%
\bibitem [{\citenamefont {Hara}\ and\ \citenamefont
  {Iwasaki}(1979)}]{hara1979}%
  \BibitemOpen
  \bibfield  {author} {\bibinfo {author} {\bibfnamefont {K.}~\bibnamefont
  {Hara}}\ and\ \bibinfo {author} {\bibfnamefont {S.}~\bibnamefont {Iwasaki}},\
  }\bibfield  {title} {\bibinfo {title} {On the quantum number projection:
  ({{I}}). {{General}} theory},\ }\href
  {https://doi.org/10.1016/0375-9474(79)90094-0} {\bibfield  {journal}
  {\bibinfo  {journal} {Nucl. Phys. A}\ }\textbf {\bibinfo {volume} {332}},\
  \bibinfo {pages} {61} (\bibinfo {year} {1979})}\BibitemShut {NoStop}%
\bibitem [{\citenamefont {Onishi}\ and\ \citenamefont
  {Yoshida}(1966)}]{onishi1966}%
  \BibitemOpen
  \bibfield  {author} {\bibinfo {author} {\bibfnamefont {N.}~\bibnamefont
  {Onishi}}\ and\ \bibinfo {author} {\bibfnamefont {S.}~\bibnamefont
  {Yoshida}},\ }\bibfield  {title} {\bibinfo {title} {Generator coordinate
  method applied to nuclei in the transition region},\ }\href
  {https://doi.org/10.1016/0029-5582(66)90096-4} {\bibfield  {journal}
  {\bibinfo  {journal} {Nucl. Phys.}\ }\textbf {\bibinfo {volume} {80}},\
  \bibinfo {pages} {367} (\bibinfo {year} {1966})}\BibitemShut {NoStop}%
\bibitem [{\citenamefont {Lieb}(1968)}]{lieb1968}%
  \BibitemOpen
  \bibfield  {author} {\bibinfo {author} {\bibfnamefont {E.~H.}\ \bibnamefont
  {Lieb}},\ }\bibfield  {title} {\bibinfo {title} {A theorem on {Pfaffians}},\
  }\href {https://doi.org/10.1016/S0021-9800(68)80078-X} {\bibfield  {journal}
  {\bibinfo  {journal} {Journal of Combinatorial Theory}\ }\textbf {\bibinfo
  {volume} {5}},\ \bibinfo {pages} {313} (\bibinfo {year} {1968})}\BibitemShut
  {NoStop}%
\bibitem [{\citenamefont {Caianiello}(1973)}]{caianiello1973}%
  \BibitemOpen
  \bibfield  {author} {\bibinfo {author} {\bibfnamefont {E.~R.}\ \bibnamefont
  {Caianiello}},\ }\href@noop {} {\emph {\bibinfo {title} {Combinatorics and
  renormalization in quantum field theory}}},\ Frontiers in physics\ (\bibinfo
  {publisher} {W.A. Benjamin},\ \bibinfo {address} {Reading, Mass},\ \bibinfo
  {year} {1973})\BibitemShut {NoStop}%
\bibitem [{\citenamefont {Terhal}\ and\ \citenamefont
  {DiVincenzo}(2002)}]{terhal2002}%
  \BibitemOpen
  \bibfield  {author} {\bibinfo {author} {\bibfnamefont {B.~M.}\ \bibnamefont
  {Terhal}}\ and\ \bibinfo {author} {\bibfnamefont {D.~P.}\ \bibnamefont
  {DiVincenzo}},\ }\bibfield  {title} {\bibinfo {title} {Classical simulation
  of noninteracting-fermion quantum circuits},\ }\href
  {https://doi.org/10.1103/PhysRevA.65.032325} {\bibfield  {journal} {\bibinfo
  {journal} {Phys. Rev. A}\ }\textbf {\bibinfo {volume} {65}},\ \bibinfo
  {pages} {032325} (\bibinfo {year} {2002})}\BibitemShut {NoStop}%
\bibitem [{\citenamefont {Hairer}\ \emph {et~al.}(1993)\citenamefont {Hairer},
  \citenamefont {Wanner},\ and\ \citenamefont {Norsett}}]{hairer1993}%
  \BibitemOpen
  \bibfield  {author} {\bibinfo {author} {\bibfnamefont {E.}~\bibnamefont
  {Hairer}}, \bibinfo {author} {\bibfnamefont {G.}~\bibnamefont {Wanner}},\
  and\ \bibinfo {author} {\bibfnamefont {S.~P.}\ \bibnamefont {Norsett}},\
  }\href {https://doi.org/10.1007/978-3-540-78862-1} {\emph {\bibinfo {title}
  {Solving {{Ordinary Differential Equations I}}}}},\ \bibinfo {series}
  {Springer {{Series}} in {{Computational Mathematics}}}, Vol.~\bibinfo
  {volume} {8}\ (\bibinfo  {publisher} {{Springer}},\ \bibinfo {address}
  {{Berlin, Heidelberg}},\ \bibinfo {year} {1993})\BibitemShut {NoStop}%
\bibitem [{zenodo()}]{zenodo}%
  \BibitemOpen
  \href@noop {} {}\bibinfo {note} {E. Mascot, T. Hodge, D. Crawford, J. Bedow, D. K. Morr,
and S. Rachel, Data repository accompanying ``Many-Body Majorana Braiding without an Exponential Hilbert space'' (2023), \href{https://zenodo.org/records/10005837}{10.5281/zenodo.10005836}.}\BibitemShut
  {Stop}%
\end{thebibliography}
%

\end{document}